\documentstyle[aps,epsfig,prl]{revtex}
\begin{document}
\title{An echo of an exciting light pulse  in quantum wells}
\author{I. G. Lang, L. I. Korovin}
\address{A. F. Ioffe Physical-Technical Institute, Russian
Academy of Sciences, 194021 St. Petersburg, Russia}
\author{D. A. Contreras-Solorio, S. T. Pavlov\cite{byline1}}
\address {Escuela de Fisica de la UAZ, Apartado Postal C-580,
98060 Zacatecas, Zac., Mexico} \twocolumn[\hsize\textwidth\columnwidth\hsize\csname
@twocolumnfalse\endcsname
\date{\today}
\maketitle \widetext
\begin{abstract}
\begin{center}
\parbox{6in}
{The non-sinusoidal character oscillations appear in the transmitted, reflected and
absorbed light fluxes when light pulses irradiate a semiconductor quantum well (QW),
containing  a large number of the equidistant energy levels of electronic
excitations. A damping echo of the exciting pulse appears through the time intervals
$2\pi\hbar/\Delta E$ in the case of the very short light pulses $\gamma_l^{-1}\ll
\hbar/\Delta E$.}
\end{center}
\end{abstract}

] \narrowtext

A row of theoretical and experimental investigations is devoted
to the elaborate study of the electronic excitations  in the bulk
crystals and semiconductor QWs with the help of the time resolved
scattering (TRS), because just the existence of the discrete
energy levels determines the most interesting results obtainable
by the TRS. It is well known that a couple of the closely
disposed energy levels demonstrate a new effect: The sinusoidal
beatings appear in reflected and transited pulses on a frequency
corresponding to the energy distance between the energy levels
\cite{1}.

We investigate theoretically below reflection and absorption of
light pulses, irradiating semiconductor QWs, having a large number
of equidistant energy levels. Let us suppose that a QW with a
system of equidistant discrete energy levels is irradiated by a
light pulse with a carrier frequency $\omega_l$. Let us suppose
that the frequency $\omega_l$ is in the resonance with one of
energy levels.

Then two variants are possible: first, one may neglect influence
of the rest energy levels; second, one have to take into account
a few neighbor energy levels. The first variant is realized, for
instance, for a symmetrical pulse under condition
$\gamma_l<<\Delta \omega$, where $\Delta \omega$ is the distance
between the neighbor levels, $gamma_l$ is the parameter
characterizing the pulse duration.  It is shown \cite{2} that
under condition $\gamma_r \geq \gamma$ , where $\gamma_r
(\gamma)$ is the radiative (non-radiative) broadening of an
electronic excitation, the transmitting pulse profile changes
drastically. The lifetimes of the electron-hole pairs (EHP) in a
QW in a quantizing magnetic field have been calculated in
\cite{3,4}. The second variant with many equidistant energy
levels has been considered in \cite{3,5}, where a ladder-like
structure of a reflected and transmitted pulses has been
predicted.

We consider here both symmetrical and asymmetrical pulses,
demonstrate the results for the different
 $\gamma_l$ and show
that very short pulses create "an echo" in reflected and
transmitted light.

  Let us consider a
QW with a system of equidistant energy levels
\begin{equation}
\label{1}
\omega_{\rho}=\rho \Omega_\mu, \qquad \rho =0,1,2...
\end{equation}
 The real
significance of the designation  $\Omega_\mu$ will be given below.
We believe that the QW's width is much smaller than the light
wavelength  $d<<c/(n\omega_l),$ where $n$ is the refraction index
outside the QW. The excitation $\rho$ is characterized by the
radiative $\gamma_{r\rho}$ and non-radiative broadenings.

We suppose that the light pulse incidents perpendicular the QW's
plane $xy$ from the left (from the negative $z$) and its electric
field is
\begin{eqnarray}
\label{2} \bf {E}_0(z,t)=E_0\bf
{e}_lexp(-i\omega_{l}p)\{\Theta(p)exp(-\gamma_{l1}p/2)\nonumber\\+
[1-\Theta(p)]exp(\gamma_{l2}p/2)\}+ c. c.,
\end{eqnarray}
where $E_0$ is the real amplitude, ${\bf e}_l$ is the circular
polarization vector, $p=zn/c, \Theta(p)$ is the Haeviside
function. We have for a symmetrical pulse
\begin{equation}
\label{3}
\gamma_{l1}=\gamma_{l2}=\gamma_l,
\end{equation}
and for an pulse with a sharp front
\begin{equation}
\label{4} \gamma_{l1}=\gamma{l},\quad \gamma_{l2}\to \infty.
\end{equation}
 The electric fields on the left (right) of the QW and
corresponding time-dependent energy fluxes are calculated
according to the methods of \cite{2,3,4,5,6}. We demonstrate some
results obtained for the following parameters
\begin{equation}
\label{5}
\gamma_{r\rho}=\gamma_r,\quad \gamma_\rho=\gamma,
\quad\gamma_r<<\gamma<<\gamma_l,\quad \gamma<<\Omega_\mu,
\end{equation}
and for an arbitrary interrelation of $\gamma_l$ and $\Omega_\mu$.
Let us consider the energy fluxes under conditions
\begin{equation}
\label{6} p>>\gamma_l^{-1},\quad s>>\gamma_l^{-1},\quad
p<<\gamma_r^{-1}, \quad s<<\gamma_r^{-1},
\end{equation}
where  $s=t+zn/c$ is the variable for the reflected energy flux.
Then the contributions into the electric fields, containing the
factors $exp(-\gamma_{l}p/2)$ or или $exp(-\gamma_{l}p/2)$,
become negligible . We obtain, that only those contributions in
induced fields are essential which are proportional to
$exp(-\gamma p/2)$ or $exp(-\gamma s/2)$. The existence of such
contributions is proved in many investigations. A picture of
induced fields outside a QW is symmetrical, since only narrow QWs
are considered. It follows from this fact that the transmitted
and reflected fluxes are equal in absolute values and the
absorbed flux equals to the doubled transmitted (or reflected)
flux with the opposite sign. A negative absorption means that the
QW gives back a stored energy, irradiating it symmetrically by
two fluxes on the left and on the right.

 The reflected energy flux results in
\begin{equation}
\label{7}
{\cal R}(s)=4(\gamma_r/\gamma_l)^2e^{-\gamma s}Y_{\Omega_l,G_l}(S),
\end{equation}
where ${\cal R}(s)$ is the module of the reflected flux  in units
$cE_0^2/(2\pi n)$, $Y_{\Omega_l,G_l}$ is the dimensionless
periodical function of the  variable $S=\Omega_\mu s$ with the
period  $2\pi$,  depending from the parameters
\begin{equation}
\label{8}
\Omega_l=\omega_l/\Omega_\mu ,\qquad G_l=\gamma_l/\Omega_l.
\end{equation}
The factor $exp(-\gamma s)$ determines damping of the reflected
flux.The function $Y$ is periodical on $s$ with the period $2\pi
/\Omega_\mu$. These oscillations are not sinusoidal.

We calculated the functions $Y_{|Omega_l,G_l}(S)$ and showed, that
the results are essentially dependent on  the parameter $G_l$.
For $G_l<<1$ the periodical vibrations of the intensities of the
reflected and transmitted energy fluxes are very small in
amplitudes. The values $\Omega_l=0,1,2...$ correspond to the
resonance of the frequency $\omega_l$ with one of the energy
levels. A small detuning of the frequency $\omega_l$ results into
the drastic dropping of the energy fluxes.

 The function
$Y_{\Omega_l,G_l}^{sym}(S)$ is represented in Figs. 1, 2 inside
of one period.

The upper index means a symmetrical exciting pulse. The function
$Y_{\Omega_l,G_l}^{sym}(S)$ is symmetrical relatively the
substitution  $S$ by на $2\pi -S$. Fig. 1 corresponds  to the
value $G_l$ and to the set  $\Omega_l =0;0.1;0.5;1;1.5;2$. The
vibration amplitude in Fig.1 reaches the value 1.5-2.  For
$\Omega_l=0.5$ and  $\Omega_l =1.5$, i. e. in the case when the
frequency is disposed between the levels  $n=0$ and и $n=1$ and
between the levels  $n=1$ and  $n=2$, respectively, the curves
touch the abscissas axis in the point $S=\pi.$

Fig. 2 demonstrates the echo of the exciting symmetrical pulse.
It corresponds to the large value  $G_l=5$ and to the set
$\Omega_l=0;0.1;0.5;1$. The function  $Y^{sym}(S)$

in the points $s=0$ и $s=2\pi$ increases sharply in comparison to
corresponding values in the Fig. 1, reaching 30, but they are
very small in the interval $s>>G_l^{-1},\quad(2\pi-S)>>G_l^{-1}$.
Thus, the periodical function $Y^{sym}(S)$ represents a sequence
of short pulses, the duration of each is of the order $G_l^{-1}$
and disposed with the intervals  $2\pi$. Applying Eq. (7), one
finds, that at $\gamma_l>>\Omega_\mu$ some echo of the exciting
pulse has to be seen in the reflected energy flux, damping as
$exp(-\gamma s)$ with the interval $2\pi/\Omega_\mu$. Such echo
must be also in the transmitted energy flux. Our results show
that in the region of values of $\Omega_l$ from 0 up to several
units,  the pulse replica form repeats. In a resonance $\omega_l$
with one of upper levels  with index $\rho_0>>1$ the form of the
repeating pulses (echo) coincides with the exciting pulse form.
The  repeating pulses contain the small factor
\begin{equation}
\label{9} \delta = \pi^2(\gamma_r/\Omega_\mu)^2exp(- \gamma s).
\end{equation}

The theory is applicable for the narrow QWs in a strong magnetic
field perpendicular to the QW's plane $xy$. Then the equidistant
levels correspond to the EHP with different Landau levels and
fixed size quantized energy levels of electrons (holes).
$\Omega_\mu=|e|H/\mu c$ is the cyclotron frequency, corresponding
to the reduced mass $\mu=m_em_h/(m_e+m_h)$, where $m_e(m_h)$ is
the electron (hole) effective mass. The energy level equidistance
is small if
\begin{equation}
\label{10} d<<a, \qquad a_H<<a,
\end{equation}
where $a$ is the radius of the Wannier=Mott exciton in a zero
magnetic field , $a_H=(c\hbar/(|e|H))^{1/2}$ is the magnetic
length, i. e. for the narrow QWs and strong magnetic fields.
 We suppose that the band's
non-parabolicity is small inside the vicinity of the band's
extremums.
\subsection*{Acknowledgements}
        S.T.P thanks the Zacatecas University and the National
Council of Science and Technology (CONACyT) of Mexico for the
financial support and hospitality. D.A.C.S. thanks CONACyT
(27736-E) for the financial support. This work has been partially
supported by the Russian Foundation for Basic Research and by the
Program "Solid State Nanostructures Physics".

\begin{figure} \caption{ The function
$Y^{sim}_{\Omega_l, G_l}(S)$, corresponding to the periodical factor in the value of
the reflected energy flux when a QW is irradiated by the symmetrical light pulse. The
pulse duration $\gamma_l^{-1}$= $\Omega_\mu^{-1}$.}
\end{figure}

\begin{figure} \caption{ Same as Fig.1 for the
very small value $\gamma_l^{-1}$, when the  echo of the exciting pulse appears.}
\end{figure}

\end{document}